\documentclass[letterpaper,12pt]{article}

\usepackage{graphicx,cite}
\usepackage{amsmath}
\usepackage{amsfonts}
\usepackage{amssymb}
\usepackage{rotating}

\usepackage{braket}

\usepackage[T1]{fontenc}

 \usepackage{times}

 \newcommand{\eV}{\ensuremath{\mathrm{eV}}}

\begin{document}
\title{The flavor-blind principle: A symmetrical approach to the Gatto-Sartori-Tonin relation} 

\author{U. J. Salda\~na-Salazar\footnote{E-mail: \texttt{ulisesjesus@protonmail.ch}}
\\
\vspace{.1cm} \\
\small{Departamento de F\'isica, Centro de Investigaci\'on y de Estudios Avanzados del IPN,}\\
\small{Apdo.~Postal 14-740, 07000, M\'exico D.F., M\'exico.} \\}

\maketitle
	\begin{abstract}
	 	We perform a systematic study of the generic Gatto-Sartori-Tonin relation, 
	 	$\tan^2\theta_{ij}= m_i/m_j$.  
		This study of fermion mixing phenomena leads us to the necessary conditions that are
		needed in order to obtain it without any approximation. 	 	
	 	We begin by considering two Dirac fermion families. 
	 	By means of the hierarchy in the masses, it is found
	 	that a sufficient
	 	and necessary condition is to have a normal matrix with $m_{11}=0$.
	 	This matrix can be decomposed into two different linearly independent contributions. 
	 	The origin for such
	 	two independent contributions can be naturally explained by what we shall call 
	 	the flavor-blind principle. 
	 	This principle states that Yukawa couplings shall be either flavor-blind
	 	or decomposed into several sets obeying distinct permutation symmetries. 
		In general, it is shown that the symmetry properties of the introduced
	 	set of Yukawa matrices follow
	 	for $n$ fermion families the unique sequential breaking $S_{nL}\otimes S_{nR} \rightarrow S_{(n-1)L}\otimes S_{(n-1)R} \rightarrow \cdots 
	\rightarrow S_{2L}\otimes S_{2R} \rightarrow S_{2A}$.
	 	The particular case of three fermion families
	 	explains why the four mass ratios parametrization that we recently proposed
		can be used even in the case of no hierarchical masses. 	 	 
	\end{abstract}

	\section{Introduction}
	Family mixing in both the quark and lepton sectors is nowadays a well established fact.
	It provides us with a description on how flavor transitions are produced when
	fermions interact with the charged bosons and also, in the case of neutrinos, through their 
	propagation. 
    The rate of these transitions are measured by the moduli of the mixing matrix elements
    whereas the amount of Charge-Parity (CP) violation  by the corresponding Jarlskog invariant. 
	The Standard Model of the electroweak interactions plus massive neutrinos,
	successfully describes this picture with
	ten physical parameters \textit{per} sector (in the case of Majorana neutrinos, two more phases). 
	Nevertheless, since the late sixties of the past century, 
	the possibility of having relations among
	the mixing angles and fermion mass ratios was realized.
	The pioneering work of Gatto, Sartori, and Tonin
	showed a relation satisfying, with good accuracy, 
	the experimental value of the Cabibbo angle~\cite{Gatto:1968ss}
	\begin{eqnarray}
		\tan^2 \theta_C \approx {\frac{m_d}{m_s}}.
	\end{eqnarray}
	At present, a more complete relation with greater accuracy includes the contribution coming from the up-quark sector and a small correction through the denominator~\cite{Lehmann:1995br, Hollik:2014jda}
	\begin{eqnarray}
		\tan^2\theta_C \approx {\frac{\hat{m}_d + \hat{m}_u}{(1+\hat{m}_d)(1+\hat{m}_u)}},
	\end{eqnarray}
	with $\hat{m}_d = m_d/m_s$ and $\hat{m}_u = m_u / m_c$.
	From now on, we will refer to it, in its generic form which only encompasses a single fermion species, $\tan^2\theta_{ij} = m_i/m_j$, 
	as the Gatto-Sartori-Tonin (GST) relation, 
	for a review on this topic 
	please refer to~\cite{Fritzsch:1999ee}. It has been already recognized by different authors 
	that not only the observed hierarchy in the quark masses~\cite{Fritzsch:1986sn, Hall:1993ni, Xing:1996hi,
	Rasin:1997pn, Rasin:1998je, Chkareuli:1998sa, Fritzsch:1999rb, Fritzsch:1999ee, Chkareuli:2001dq,Morisi:2011pt,
	Hollik:2014jda}
	\begin{eqnarray}
		m_t(M_Z) \gg m_c(M_Z) \gg m_u(M_Z), \\
		m_b(M_Z) \gg m_s(M_Z) \gg m_d(M_Z),
	\end{eqnarray}
	can give access to desired GST relations, but also the charged leptons hierarchy plus the
	possible milder one in the neutrino masses~\cite{Morisi:2009sc,Morisi:2011pt,Xing:2012zv, Hollik:2014jda} 
	\begin{eqnarray}
		m_\tau(M_Z) \gg m_\mu(M_Z) \gg m_e(M_Z), \\
		m^2_{\nu 3}(M_Z) \gg m^2_{\nu 2}(M_Z) \gg m^2_{\nu1}(M_Z).
	\end{eqnarray}
	However, only until recently, a procedure to build
	a mixing parametrization in terms of only the four 
	independent mass ratios of the corresponding  fermion sector was fully achieved~\cite{Hollik:2014jda}. 
	This new parametrization makes one to wonder about the possible 
	physical meaning of the GST relation and the necessary conditions to find it in an exact way. 
	Moreover, the procedure allowed the computation of the neutrino masses predicting 
	the mass values~\cite{Hollik:2014jda}
\begin{align*}
m_{\nu1} &= ( 0.0041 \pm 0.0015 )\,\eV, \\
m_{\nu2} &= ( 0.0096 \pm 0.0005 )\,\eV, \\
m_{\nu3} &= ( 0.050 \pm 0.001 )\,\eV.
\end{align*}	
	From them, the study of the leptonic mixing was made possible and was
	found to be in excellent agreement to the most recent global fits~\cite{Hollik:2014jda}. Nevertheless,
	the existing mild hierarchy in the neutrino masses cannot wholly justify this procedure
	which was obtained from the strong hierarchy in the quark masses. Therefore, the only reason
	to explain such an excellent agreement in the leptonic mixing despite the mild hierarchy in the neutrino masses
	is to disentangle the origin of 
	the hierarchical nature from the dynamics responsible for fermion mixing, see for example~\cite{Knapen:2015hia}. In the following, we leave aside the issue of hierarchical
	masses and only focus on what is being responsible for fermion mixing.
	
	Yukawa interactions between the different fermion families
	with the Higgs field are only requested to be renormalizable and gauge invariant.
	As a consequence, Yukawa couplings appear as a set of arbitrary complex parameters.
	But, could this set of Yukawa couplings	follow some principle? 
	In this work we try to answer this question.
	We take  the study of the GST relation
	in the two family case
	as our starting point. We find that a Yukawa matrix should be
	normal
	with its first diagonal element equal to zero, $y_{11} =0$, to be necessary and sufficient in order
	to be diagonalized by the GST relation. This in return permits us to distinguish two
	different hierarchical structures, $y_{22} \gg y_{ij}$; 
	their symmetric or antisymmetric properties thus pointing out
	to a possible symmetrical origin. This basis is commonly known as heavy or
	hierarchical basis, as suggested by its structure.
	 Nonetheless, within this initial basis, we
	 note that the dominant contribution is invariant under a
	$U(1)^3$ global symmetry whereas the small contribution breaks it. 
	An appropriate weak basis transformation can put them into equal footing.\footnote{A 
	weak basis transformation is such that the charged current remains diagonal after it.}
	In the new basis, it may be read (what we shall call the flavor-blind principle): 
	\textit{Yukawa couplings shall be either flavor-blind
	 	or decomposed into several sets obeying distinct permutation symmetries}. This statement for the three family case was first initiated but partially by Harari, Haut, and 
	Weyers~\cite{Harari:1978yi}
	and later reconsidered with different approaches by other authors~
	\cite{Kaus:1988tq,Lavoura:1989dx,Fritzsch:1989qm,Tanimoto:1989qh,Kaus:1990ij,Babu:1990fr,
	Fritzsch:1994yx,Lehmann:1995br,Mondragon:1998gy,Barranco:2010we}.
	The main idea was to consider, as onset, the so called
	 ``democratic form'' which is invariant under permutations of the columns and/or
	rows, and described by the symmetric group $S_{3L}\otimes S_{3R}$. Afterwards, 
	the addition of further terms  would explicitly break it into various ways. 
	Here we show that the flavor-blind principle gives, on its own, all the necessary information to
	build up 
	the Yukawa matrices; putting aside, the initial arbitrariness with which they were introduced. 
	Our approach differs from previous ones in several aspects:\footnote{Recall that we
	do not intend here to give an explanation to the hierarchical nature of fermion masses,
	but rather show how mixing phenomena are driven by the flavor-blind principle. For example,
	the formalism to be developed here could complement recent works treating the hierarchy in 
	the masses~\cite{Ibarra:2014fla,Altmannshofer:2014qha,Ibarra:2014pfa,Knapen:2015hia}.}
	\begin{enumerate}
		\item We assume as our starting point that fermion mixing is described by the GST relation.
		\item From it, we find two necessary and sufficient
		 mathematical conditions to be satisfied by a Yukawa matrix: normality and its $1-1$ 		
		 element equal to zero.
		\item This mathematical condition can be reexpressed as a physical statement, which
		we call the flavor-blind principle:
		Yukawa couplings shall be either flavor-blind
	 	or decomposed into several sets obeying distinct permutation symmetries.
		\item The main purpose pursued in earlier works~\cite{Harari:1978yi,Kaus:1988tq,Lavoura:1989dx,Fritzsch:1989qm,Tanimoto:1989qh,Kaus:1990ij,
		Babu:1990fr,
	Fritzsch:1994yx,Lehmann:1995br,Mondragon:1998gy,Barranco:2010we}
		 was to build mass matrices with some of its elements being equal to zero, and therefore justify how texture zeros arise. The symmetry was explicitly broken from $S_{3L} \otimes S_{3R}$ to either $S_{2L} \otimes S_{2R}$~\cite{Fritzsch:1989qm,Kaus:1990ij,Fritzsch:1994yx,Lehmann:1995br
		 ,Mondragon:1998gy,Barranco:2010we} or completely~\cite{Harari:1978yi,Kaus:1988tq,Lavoura:1989dx,Tanimoto:1989qh,Babu:1990fr}. In the former case, it would afterwards be broken completely by terms either being $S_3^{\text{diag}}$ symmetrical~\cite{Barranco:2010we,Mondragon:1998gy,Lehmann:1995br,Fritzsch:1989qm} or without any symmetry at all~\cite{Kaus:1990ij,Fritzsch:1994yx}. These different ways seem to follow only a subjective choice. Here we want to stress that this is not the case and the breaking pattern follows a unique ordering
		\begin{eqnarray}
			S_{3L}\otimes  S_{3R} \rightarrow S_{2L} \otimes S_{2R} \rightarrow S_{2A}.
		\end{eqnarray}
		\item The latter symmetry breaking pattern is partially equivalent to the
		minimal breaking of the flavor symmetry group which appears as an accidental global
		symmetry when Yukawa interactions are turned off
		\begin{eqnarray}
			U(3)^6  \rightarrow U(2)^6 \rightarrow U(1)^6 \rightarrow U(1)_B \otimes U(1)_L,
		\end{eqnarray}
		but it is not completely equivalent as it lacks an explanation to 
		the hierarchical nature of fermion masses.
		\item In general, the flavor-blind principle applied to the $n$ family case implies the symmetry
		 breaking 	pattern
		\begin{eqnarray}
			S_{nL}\otimes  S_{nR} \rightarrow 
			S_{(n-1)L}\otimes  S_{(n-1)R} \rightarrow \cdots \rightarrow
			S_{3L}\otimes  S_{3R} \rightarrow 
			S_{2L} \otimes S_{2R} \rightarrow S_{2A}.
		\end{eqnarray}
	\end{enumerate}
	
	This paper is organized in the following way. In Section 2, we study the two fermion 
	family case. We	begin by finding
	 the necessary conditions a mass matrix should satisfy to be diagonalized by the GST relation.
	 Afterwards, we study the implications of 
	 demanding one of the mass matrix elements to be equal to zero.
	In Section 3, we find a symmetrical origin which not only works in the two family case
	but also for the three family one. Finally, we conclude. 
	In the Appendix~\ref{App:2qho} we remark the universal nature of the GST relation by
	decoupling two quantum harmonic oscillators with equal frequencies but different masses; 
	while  in Appendix~\ref{app:nFamiliesCase} we apply the flavor-blind principle to the $n$ family case
	and show why the last step of the symmetry breaking pattern is $S_{2A}$ and not
	$S_{2S}$.

	\section{Two fermion families}
	The GST relation is a universal relation. By this we mean that it appears in different kinds of
	physical phenomena, as for example, when decoupling two coupled
	quantum harmonic oscillators with 
	different masses but equal frequencies, see Appendix~\ref{App:2qho}. 
	In general, the relation may be derived from the quadratic equation 	
	\begin{align} \label{quadraticTheta}
		\tan^2\theta + \frac{m_2 - m_1}{\sqrt{m_1 m_2}} \tan\theta -1 = 0,
	\end{align}
	which has two solutions:
	\begin{align}
	\tan\theta = \begin{cases}
    -\sqrt{\frac{m_2}{m_1}}\\
    {}\\
    + \sqrt{\frac{m_1}{m_2}}
	  \end{cases},
	\end{align}
	corresponding the latter to the GST relation.
	
	What are the necessary and sufficient conditions to obtain this relation in an	\textit{exact} 
	manner within the context of two fermion families mixing?	
	
	\subsection{Finding out the conditions}
	Following a similar procedure as the authors in~\cite{Hollik:2014jda} we can use the hierarchy in the masses 
	to study the matrix elements. The implied most general form for the mass matrices is a normal matrix
	given as\footnote{The
	 contribution
	of ${\bf m}_{11}$ in ${\bf m} {\bf m}^\dagger$, if we assume it to be much smaller than one, 
	could be neglected in the calculations~\cite{Hollik:2014jda}.
	However, we will leave out any aproximations and stick to the general case.}
	\begin{eqnarray}
		{\bf m} = \begin{pmatrix}
			\kappa |m_{12}|^2 e^{i {\delta_{11}}} & |m_{12}| e^{i \delta_{12}} \\
			|m_{12}| e^{i \delta_{21}} & |m_{22}| e^{i \delta_{22}} 
		\end{pmatrix},
	\end{eqnarray}
	where the coefficient $\kappa$ has been introduced in order to track the contributions coming from $m_{11}$.
	
	The obtention of the unitary transformation which enters in the charged current interactions 
	requires the study of	the left Hermitian product, ${\bf m} {\bf m}^\dagger$. In order to put this product
	into diagonal form we require, in general, a unitary transformation with the rotation angle satisfying
	\begin{eqnarray} \label{quadraticUnknown}
		\tan^2\theta - \left( \frac{n_{11} - n_{22}}{|n_{12}|} \right) \tan \theta - 1 = 0,
	\end{eqnarray}
	where we have denoted the left Hermitian product by ${\bf n} = {\bf m} {\bf m}^\dagger$ 
	and its matrix elements are
	\begin{align}
		n_{11} = \kappa^2|m_{12}|^4 + |m_{12}|^2, \quad\quad\quad\quad
		n_{22} = |m_{12}|^2 + |m_{22}|^2, \\
		n_{12} = \kappa|m_{12}|^3 e^{i(\delta_{11} - \delta_{21})}+ |m_{12}||m_{22}|e^{i(\delta_{12} - \delta_{22})}. 
		\hspace{1cm}
	\end{align}
	
	On the other hand, if what we want is to reproduce the GST relation, then Eqs.~(\ref{quadraticTheta}) 
	and~(\ref{quadraticUnknown}) imply our first condition which is sufficient to produce the desired
	relation
	\begin{eqnarray} \label{suff-cond}
		 \frac{n_{11} - n_{22}}{|n_{12}|} = \frac{m_1 - m_2}{\sqrt{m_1 m_2}}.
	\end{eqnarray}
	In terms of matrix elements it reads
	\begin{eqnarray}
		\frac{\kappa^2|m_{12}|^4 - |m_{22}|^2}{\sqrt{\kappa^2|m_{12}|^6 + |m_{12}|^2|m_{22}|^2 + 2\kappa |m_{12}|^4|m_{22}|\cos(\delta_{11} + \delta_{22} - (\delta_{12} + \delta_{21}))}} = \frac{m_1 - m_2}{\sqrt{m_1 m_2}}.
	\end{eqnarray}
	At this point, it is instructive to make use of the two invariants of the mass matrix
	\begin{eqnarray}
         {\text{Tr}({\bf n}) } =n_{11} + n_{22} = m_1^2 + m_2^2, \\
         {\text{det}({\bf n}) } =n_{11}n_{22} - |n_{12}|^2 = m_1^2m_2^2,
    \end{eqnarray}
	 which in terms of the matrix elements are
	 \begin{eqnarray}
	 	\kappa^2 |m_{12}|^4 + 2|m_{12}|^2 + |m_{22}|^2 = m_1^2 + m_2^2, \\
	 	|m_{12}|^4 \left[ \kappa^2 |m_{22}|^2 - 2 \kappa |m_{22}| \cos(\delta_{11} + \delta_{22} - (\delta_{12} + \delta_{21})) + 1 \right] = m_1^2 m_2^2.
	 \end{eqnarray}
	 
	 	A first thing  to notice here is that we have an undesired dependence 
	 	on a  linear combination of phases.	Recall that the GST relation only depends on the masses
	 	and we want to have this same dependence on the moduli of the mass matrix elements, 
	 	$|m_{ij}| = |m_{ij}|(m_1,m_2)$.\footnote{A more general treatment could allow the possible dependence to complex phases
	 	\begin{eqnarray}
	 		|m_{ij}| = |m_{ij}|(m_1,m_2, \delta_{11}, \delta_{12}, \delta_{21}, \delta_{22}).
		\end{eqnarray}	 	 
		However, we will assume from this point onwards that the moduli of the matrix elements do not have this dependence.} 
	 	This in return will give us Yukawa couplings whose magnitude will only depend on the masses. 
	 	To get rid off this kind of phase dependence then we
	 	require to either have $m_{11}$ or $m_{22}$ equal to zero. In fact, the only viable option
	 	consistent with the hierarchical structure of ${\bf m} {\bf m}^\dagger$ is $m_{11} = 0$.
	 	So we set $\kappa = 0$ and obtain a solution consistent with Eq.~(\ref{suff-cond}),
	 	\begin{eqnarray}
	 		|m_{12}| = \sqrt{m_1 m_2}, \quad\quad\quad\quad
	 		|m_{22}| = m_2 - m_1	.
		\end{eqnarray}	 	 
	 	
	 	We then conclude that
	 	setting $\kappa = 0$ is necessary if we want the magnitude of the Yukawa 
	 	couplings to only depend on the masses
	 	and not on phases coming from the complex matrix elements.
	 	
	 	Hence, we find the form of a complex matrix sufficient to reproduce the GST relation by
		\begin{eqnarray}
	 		{\bf m} = \begin{pmatrix}
	 			0 & \sqrt{m_1 m_2} e^{i \delta_{12}} \\
			\sqrt{m_1 m_2} e^{i \delta_{21}} & (m_2-m_1) e^{i \delta_{22}} 
	 		\end{pmatrix}.
	 	\end{eqnarray}	 
	 	
	 	\subsection{Implications of $m_{11} = 0$}
	 	We define the singular value decomposition of the mass matrix as
	 	\begin{eqnarray}
	 		{\bf m} = {\bf L}^\dagger {\bf \Sigma} {\bf R}
	 	\end{eqnarray}
	 	where ${\bf \Sigma} = {\text{diag}}(m_1,m_2)$ and ${\bf L}$ and ${\bf R}$ are the unitary transformations 
	 	which bring the left and right
	 	Hermitian products, ${\bf m} {\bf m}^\dagger$ and ${\bf m}^\dagger {\bf m}$, 
	 	to diagonal form. 
	 	The latter equation can be written explicitly as
	 	\begin{eqnarray}
		{\bf m} = 
			m_1 \begin{pmatrix}
				L_{11}^* R_{11} & L_{11}^* R_{12} \\   L_{12}^* R_{11} & L_{12}^* R_{12}
			\end{pmatrix} 
			+ m_2 \begin{pmatrix}
				L_{21}^* R_{21} & L_{21}^* R_{22} \\   L_{22}^* R_{21} & L_{22}^* R_{22}
			\end{pmatrix},
		\end{eqnarray}
		which plainly says to us that
		in general $m_{11}$ is different from zero,
		\begin{eqnarray}
			m_{11} = m_1 L_{11}^* R_{11} + m_2 L_{21}^* R_{21} \neq 0.
		\end{eqnarray}
		Considering the matrices ${\bf L}$ and ${\bf R}$ as $SU(2)$ transformations with their rotation angle satisfying $\tan^2 \theta = m_1/m_2$
		the latter equation transforms into
		\begin{eqnarray}
			m_{11} = \frac{m_1 m_2 \pm m_1 m_2 e^{-i(\delta_L - \delta_R)}}{m_1 + m_2},
		\end{eqnarray}
		where the choice in sign is the freedom of having one rotation clockwise and the other
		either the same or counter-clockwise, respectively. Therefore, if by means of the left and right
		unitary transformations
		we want to guarantee this relation to be zero
		then we should have $\delta_L = \delta_R$ and opposite rotations for the left- and right-handed
		fields or likewise $\delta_L = \delta_R + \pi$ with both rotations in the same direction.
		
		Our final mass matrix becomes with $\delta \equiv \delta_L$
		\begin{eqnarray}
			{\bf m} = \begin{pmatrix}
				0 & \sqrt{m_1 m_2} e^{-i\delta} \\
				-\sqrt{m_1 m_2} e^{i\delta} & m_2 - m_1
			\end{pmatrix},
		\end{eqnarray}
		which is an anti-Hermitian matrix with $\delta_{22}=0$ as implied by $\delta_L = \delta_R + \pi$. 
		
		In the three family case, it has been shown~\cite{Masina:2005hf,Masina:2006ad} that it suffices to take either
		real or purely imaginary mass matrix elements to reproduce the experimentally observed
		masses and mixings. We assume this same feature in the two family case. Hence, our phase will be constrained to take only the values
		$\delta = 0 , \pi/2, \pi,\text{ or } 3\pi/2$. In fact, the same set of values can be inferred by recalling that in the two (three) family case
		a mixing parametrization requires one (four) independent parameter (parameters) to be fully described. It is only when the number of fermion families is $n\leq 3$ that a mixing parametrization has enough fermion mass ratios to contain the number of mixing parameters~\cite{Hollik:2014jda}.
		And thus, as the two (four) mass ratios parametrization indicates, 
		phases should not play any role
		except to determine if a rotation is real or its off-diagonal elements are
		purely imaginary with a freedom of choosing it
		either clockwise or anti-clockwise~\cite{Hollik:2014jda}.
		
		Now, inquiring the nature of such Yukawa interactions one is led to think about a possible
		origin for such a matrix form.	A look into its hierarchical structure suggests a natural
		separation among its elements
		\begin{eqnarray} \label{2fam-symmoranti}
			{\bf m} = (m_2 - m_1)\begin{pmatrix}
				0 & 0 \\ 0 & 1			
			\end{pmatrix} +
				\sqrt{m_1 m_2}\begin{pmatrix}
				0 &  e^{-i\delta} \\
				-e^{i\delta} & 0
			\end{pmatrix}.			
		\end{eqnarray}		
		Note that the first term is a symmetric matrix
		and when $\delta=0\text{ or }\pi$ the second term is antisymmetric whereas symmetric 
		for $\delta=\pi/2\text{ or }3\pi/2$. Are these matrix structures dictated by some principle?  
		
		\section{A symmetrical origin}	
		Why are Yukawa interactions not flavor-blind though gauge interactions are?
		With this we mean that if $n$ fields had the same gauge interactions and gauge couplings,
		 then the strength of their 
		Yukawa couplings to the Higgs field should not distinguish the flavor.
		The most general form for this assertion
		is to introduce Yukawa couplings ${\bf y}_{ij}$ 
		which are invariant under permutations of the massless families, that is flavor-blind Yukawa 
		couplings (see Appendix~\ref{app:nFamiliesCase}).

		\subsection{Two fermion families}
		Two fermion families coupling without distinction to the Higgs field imply
		\begin{eqnarray} \label{2fam-m2}
			{\bf y}^{1 \leftrightarrow 2} = y \begin{pmatrix}
				1 & 1 \\ 1 & 1
			\end{pmatrix}.
		\end{eqnarray}
		This basis in commonly known as the democratic basis.
		Then in return, after the weak basis transformation 
		\begin{eqnarray} \label{eq:2fam-wbt}
			O_2 = \frac{1}{\sqrt{2}} \begin{pmatrix}
					1 & 1 \\
					-1 & 1
			\end{pmatrix},				
		\end{eqnarray}				
		the latter Yukawa couplings imply one massless, $m_1=0$,
		and one massive family, $m_{2} = 2y v$ with $v\equiv \braket{H}$
		the nonvanishing vacuum expectation value of the neutral component of the Higgs field. 
		This new basis is called the heavy o hierarchical basis. Here the Yukawa matrix takes
		the form
		\begin{eqnarray}
			\widetilde{\bf y}^{1 \leftrightarrow 2} = 2 y \begin{pmatrix}
			0 & 0 \\
			0 & 1
			\end{pmatrix}.
		\end{eqnarray}
		
		The matrix in Eq.~(\ref{2fam-m2}) has the symmetry
		properties of the group $S_{2L} \otimes S_{2R}$, that is invariance under the separate
		permutations of the rows and columns. At this point, it is not clear how the first family could
		acquire its mass. In fact, we have two choices: in the democratic basis we can either break $S_{2L} \otimes S_{2R}$
		to $S_{2S}$ or $S_{2A}$, that is the symmetric or antisymmetric diagonal subgroup,
		respectively. The former is discarded as it will only give mass to the first family without inducing
		any mixing phenomena. On the other hand, the latter, 
		described by the matrix term
		\begin{eqnarray}
			{\bf y}^{A} = \begin{pmatrix}
				\beta & \alpha \\
				-\alpha & -\beta
			\end{pmatrix},
		\end{eqnarray}
		will imply, after the weak basis transformation of Eq.~(\ref{eq:2fam-wbt}), 
		\begin{eqnarray}
			\widetilde{\bf y}^A = \begin{pmatrix}
				0 & \beta + \alpha \\
				\beta - \alpha & 0
			\end{pmatrix}.
		\end{eqnarray}
		In order to have the off-diagonal elements equal in magnitude, as in Eq.~(\ref{2fam-symmoranti}), we need $\alpha$ to be purely imaginary. 
		Thus,
		\begin{eqnarray}
			\widetilde{\bf y}^A = \begin{pmatrix}
				0 & |y|_A e^{-i\delta_A} \\
				|y|_A e^{i\delta_A} & 0
			\end{pmatrix},
		\end{eqnarray}
		where $|y|_A^2 = \alpha^2 + \beta^2$ and $\tan \delta_A = - \alpha/\beta$.  When
		$\delta_A = 0 \text{ or } \pi$ ($\alpha = 0$)
		our matrix is symmetric and when $\delta_A = \pi/2 \text{ or } 3\pi/2$ ($\beta = 0$) then
		our matrix becomes antisymmetric. Just what we were looking for. The need for the
		symmetry breaking pattern $S_{2L} \otimes S_{2R} \rightarrow S_{2A}$ 
		in the two family case was recognized before~\cite{Lehmann:1995br}, although 
		it was found under the \textit{ansatz} of an Hermitian mass matrix and the
		{consideration} that in some special basis one should have $y_{11} = 0$.

		Hence in the two family case, we have found a statement 
		equivalent to our unique condition for reproducing the 
		GST relation:  \textit{Yukawa couplings shall be either flavor-blind
	 	or decomposed into several sets obeying distinct permutation symmetries.}	
		 We shall call this statement the \textit{flavor-blind principle}.	In this sense, 
		 we are taking the concept of
		 flavor-blind gauge interactions to the Yukawa sector by considering Yukawa couplings which 
		 cannot distinguish massless fields with equal quantum numbers.

		\subsection{Three fermion families}
		A recent study of fermion mixing in terms of mass ratios~\cite{Hollik:2014jda}
		showed that the three by three complex mass matrix has a hierarchical
		structure given by
		\begin{eqnarray} \label{Hier-Mass-Matrix}
			{\cal M} \sim f_3\begin{pmatrix}
				0 & 0 & 0 \\
				0 & 0 & 0 \\
				0 & 0 & 1
			\end{pmatrix}
			+
			f_2 \begin{pmatrix}
				0 & 0 & 0 \\
				0 & 0 & \times \\
				0 & \times & 0
			\end{pmatrix}
			+
			f_1 \begin{pmatrix}
				0 & \times & \times \\
				\times & 0 & 0 \\
				\times & 0 & 0
			\end{pmatrix},
		\end{eqnarray}
		with $f_i = f_i (m_1, m_2, m_3)$ an homogeneous function of degree one in $m_3$ and 
		obeying the relation 
		$f_3 \gg f_2 \gg f_1$. The particular feature of~\cite{Hollik:2014jda} is that the three fermion 
		families mixing	is decomposed in the study of successive two fermion families mixing. 
		In fact, two independent questions emerge and should concern 
		us: first, what lies behind the hierarchical nature between the matrix
		terms? and second,
		how these different Yukawa matrices appear? This work gives an explanation to the
		latter inquiry. 
		 
		We wish now to apply our flavor-blind principle to the three family case to find out how it works
		in this scenario.
		Similarly, as in the two family case,
    the first term originates from the fact that three undistinguishable fermion
	families are coupling to the Higgs field
	\begin{eqnarray}
		{\cal Y}^{1\leftrightarrow 2 \leftrightarrow 3} = y\begin{pmatrix}
			1 &1 & 1 \\
			1 & 1 & 1 \\
			1 & 1 & 1
		\end{pmatrix}.
	\end{eqnarray}
	The symmetry property of this matrix is 
	$S_{3L} \otimes S_{3R}$. After the weak basis transformation
		\begin{eqnarray} \label{eq:O3}
			O_3 =  \begin{pmatrix}
					\frac{1}{\sqrt{2}} & \frac{1}{\sqrt{6}} & \frac{1}{\sqrt{3}} \\
					-\frac{1}{\sqrt{2}} & \frac{1}{\sqrt{6}} & \frac{1}{\sqrt{3}} \\
					0 & -\frac{2}{\sqrt{6}} & \frac{1}{\sqrt{3}}
			\end{pmatrix},				
		\end{eqnarray}
	only one family becomes massive, $m_3 = 3 y v$,
	while
	the other two remain massless. 
	
	In the mass basis,\footnote{Here the mass basis coincides with the heavy basis.} we have two massless fermion families. Therefore, the next set of Yukawa 
	couplings after returning to the initial democratic basis should be invariant under the permutations of the two massless families
	\begin{eqnarray}
		{\cal Y}^{1 \leftrightarrow 2} = \begin{pmatrix}
			\beta & \beta & \alpha_1 \\
			\beta & \beta & \alpha_1 \\
			\alpha_2 & \alpha_2 & \gamma
		\end{pmatrix},
	\end{eqnarray}
	where the symmetry property of this matrix is $S_{2L}\otimes S_{2R}$. Using the weak basis 
	transformation of Eq.~(\ref{eq:O3}) this Yukawa matrix transforms into
	\begin{eqnarray}
		\widetilde{\cal Y}^{1 \leftrightarrow 2} = \begin{pmatrix}
0 & 0 & 0 \\
0 &  -\frac{2}{3} ( \alpha_1 + \alpha_2 -\beta -\gamma ) &
       \frac{1}{3} \sqrt{2} (\alpha_1 - 2\alpha_2 +2 \beta -\gamma ) \\
0 & \frac{1}{3} \sqrt{2} (-2\alpha_1 + \alpha_2 +2 \beta -\gamma ) &  
       \frac{1}{3} (2 (\alpha_1 + \alpha_2 + 2\beta )+\gamma ) \\
		\end{pmatrix}.
	\end{eqnarray}
	Demanding it to reproduce the two family case, that is the structure appearing in Eq.~(\ref{2fam-symmoranti}) 
	implies the set of solutions shown in Table~\ref{table-3fam-rank2-sols}.
	
\begin{table}
\caption{Application of the flavor-blind principle, in the three family case with two massive families, gets parametrized by the Yukawa couplings $\alpha_1$, $\alpha_2$, $\beta$, and $\gamma$. The requirement of reproducing the GST relation, as in the two family case, for each possible value of $\delta= 0,\pi/2,\pi,\text{ or }3\pi/2$, implies a  different solution, as shown below.}
\label{table-3fam-rank2-sols}
\centering
\begin{tabular}{c|cccc}
\hline \hline
$\delta$ & $\alpha_1$ & $\alpha_2$ & $\beta$ & $\gamma$  \\
\hline \hline 
$0$  & 
$-\frac{{m_2}}{3v}+\frac{\sqrt{{m_2}{m_3}}}{ \sqrt{2}v}$ &
$-\frac{{m_2}}{3v}-\frac{\sqrt{{m_2} {m_3}}}{ \sqrt{2}v}$ & 
$-\frac{{m_2}}{3v}$ & $-\frac{{m_2}}{3v}$  \\
$\pi/2$ &  $-\frac{{m_2}}{3v}+i\frac{\sqrt{{m_2}{m_3}}}{3 \sqrt{2}v}$ &
$-\frac{{m_2}}{3v}+i\frac{\sqrt{{m_2} {m_3}}}{ 3\sqrt{2}v}$ & 
$-\frac{{m_2}}{3v}-\frac{i}{3v}  \sqrt{2{m_2}{m_3}}$ & 
$-\frac{{m_2}}{3v}+\frac{2i}{3v}  \sqrt{2{m_2}{m_3}}$  \\
$\pi$  & $-\frac{{m_2}}{3v}-\frac{\sqrt{{m_2}{m_3}}}{ \sqrt{2}v}$ &
$-\frac{{m_2}}{3v}+\frac{\sqrt{{m_2} {m_3}}}{ \sqrt{2}v}$ & 
$-\frac{{m_2}}{3v}$ & $-\frac{{m_2}}{3v}$  \\
$3\pi/2$ & $-\frac{{m_2}}{3v}-i\frac{\sqrt{{m_2}{m_3}}}{3 \sqrt{2}v}$ &
$-\frac{{m_2}}{3v}-i\frac{\sqrt{{m_2} {m_3}}}{ 3\sqrt{2}v}$ & 
$-\frac{{m_2}}{3v}+\frac{i}{3v}  \sqrt{2{m_2}{m_3}}$ & 
$-\frac{{m_2}}{3v}-\frac{2i}{3v}  \sqrt{2{m_2}{m_3}}$ \\
\hline
\end{tabular}
\end{table}

	We have reached the point where we have one massless family left. The following term brings the
	Yukawa couplings from $S_{2L}\otimes S_{2R}$ to its diagonal subgroup $S_{2A}$
	(see Appendix~\ref{app:nFamiliesCase}
	 for more details on why $S_{2A}$ and not its symmetric counterpart, $S_{2S}$):
	\begin{eqnarray} \label{Mass-1stfamily}
		{\cal Y}^A = \begin{pmatrix}
				\tau & i \mu & \nu_1 \\
				-i\mu & -\tau & -\nu_1 \\
				\nu_2 & -\nu_2 & 0	
		\end{pmatrix},				
	\end{eqnarray}
	which after the weak basis transformation of Eq.~(\ref{eq:O3}) turns into
	\begin{eqnarray} \label{3fermion-yA}
		\widetilde{{\cal Y}}^A = 
\begin{pmatrix}
 0 & \frac{i \mu -2{\nu_1}+\tau }{\sqrt{3}} & \sqrt{\frac{2}{3}} (i \mu +{\nu_1}+\tau ) \\
 \frac{-i \mu -2 {\nu_2}+\tau }{\sqrt{3}} & 0 & 0 \\
 \sqrt{\frac{2}{3}} (-i \mu + {\nu_2}+\tau ) & 0 & 0
\end{pmatrix}.
	\end{eqnarray}
	
	Here we need to stress the following. The solutions appearing in Table~\ref{table-3fam-rank2-sols} correspond to two massive families. After the inclusion of Eq.~(\ref{Mass-1stfamily})
	the solutions will suffer the following change
	\begin{eqnarray}
		m_2 \rightarrow \frac{\Delta_{23}^2}{m_3},
	\end{eqnarray}
	where $\Delta_{23} = \Delta_{23} (m_1, m_2, m_3)$ represents a sum of terms 
	to be rotated away; giving in the end, the total needed rotation  in the 2-3 sector
	of the corresponding fermion species ($f=\,u,\, d,\, e, \, \nu$)~\cite{Hollik:2014jda}.
	
	The arbitrary parameters appearing in Eq.~(\ref{Mass-1stfamily}) can be solved
	in terms of $\Delta_{13}$ and $\Delta_{12}$. We can have four different cases: symmetry
	or antisymmetry in both the 1-2 and 1-3 sectors,  or the two independent combinations. 
	The four solutions are shown in Table~\ref{table-3fam-rank3-sols}.
	It is not our purpose to give the explicit expressions in terms of the masses, but rather to show
	how each new matrix term can be solved in a unique way. 
	
	Let us summarize. In the three family case,
	the entirely Yukawa matrix for any fermion species
	is built as the sum of three different contributions
	\begin{eqnarray}
		{\cal Y} = {\cal Y}^{1\leftrightarrow 2 \leftrightarrow 3}
		+ {\cal Y}^{1\leftrightarrow 2} + {\cal Y}^A.
	\end{eqnarray}
	Each of them correspond to a step where a given number of fields were massless. 
	Due to this and by the application of the flavor-blind principle, each matrix have been demanded to 
	satisfy symmetry properties under
	permutations of the corresponding massless fields. Only in the last step, with a single
	massless family, the symmetry group is chosen as to produce mixing phenomena.
		
	\begin{table}
\caption{Application of the flavor-blind principle, in the three family case with three massive families, gets parametrized by the Yukawa couplings $\tau$, $\mu$, $\nu_1$, and $\nu_2$. The requirement of reproducing the GST relation, as in the two family case, for each possible value of $\delta= 0,\pi/2,\pi,\text{ or }3\pi/2$, implies a  different solution, as shown below. In the left column we have denoted S and A to the two possibilities of having
the off diagonal terms as either symmetric or antisymmetric, respectively.}
\label{table-3fam-rank3-sols}
\centering
\begin{tabular}{c|cccc}
\hline \hline
(1-2)-(1-3) & $\tau$ & $\mu$ & $\nu_1$ & $\nu_2$  \\
\hline \hline 
S-S  & 
$\frac{1}{3}(\sqrt{3}\Delta_{12} + \sqrt{6}\Delta_{13})$ &
$0$ & 
$\frac{1}{\sqrt{6}}(-\sqrt{2}\Delta_{12} + \Delta_{13})$  &
$-\frac{1}{\sqrt{6}}(\sqrt{2}\Delta_{12} - \Delta_{13})$\\
A-A &  
$0$ &
$-\frac{i}{3}(\sqrt{3}\Delta_{12} + \sqrt{6}\Delta_{13})$  & 
$\frac{1}{\sqrt{6}}(-\sqrt{2}\Delta_{12} + \Delta_{13})$  &
$\frac{1}{\sqrt{6}}(\sqrt{2}\Delta_{12} - \Delta_{13})$\\
S-A  &
$\frac{1}{\sqrt{3}}\Delta_{12}$  &
$-i\sqrt{\frac{2}{3}}\Delta_{13}$ & 
$\frac{1}{\sqrt{6}}(-\sqrt{2}\Delta_{12} + \Delta_{13})$  &
$-\frac{1}{\sqrt{6}}(\sqrt{2}\Delta_{12} + \Delta_{13})$  \\
A-S & 
$\sqrt{\frac{2}{3}}\Delta_{13}$  &
$-i\frac{1}{\sqrt{3}}\Delta_{12}$ & 
$\frac{1}{\sqrt{6}}(-\sqrt{2}\Delta_{12} + \Delta_{13})$  &
$\frac{1}{\sqrt{6}}(\sqrt{2}\Delta_{12} + \Delta_{13})$ \\
\hline
\end{tabular}
\end{table}

\section{Conclusions}
We have found, in the two family case, the minimal form the Yukawa matrices should satisfy in order to lead exactly to a  Gatto-Sartori-Tonin (GST) like fomula for an individual fermion species, $\tan^2\theta_{ij}^f = m_{f,i}/m_{f,j}$ ($f=\,u,\, d,\, e, \, \nu$). Furthermore, we have also found that this form can be obtained from demanding the Yukawa matrices to possess symmetry properties under certain permutations of its rows and/or columns. This symmetrical origin has been expressed by the following statement which we have called the flavor-blind principle: 
\textit{Yukawa couplings shall be either flavor-blind
	 	or decomposed into several sets obeying distinct permutation symmetries.} 
Fermion mixing and the hierarchical massive nature of fermions can be studied as two independent features (see for example~\cite{Knapen:2015hia}): on one hand, some dynamics giving hierarchical masses and on the other, some mechanism producing the mixing in a given ordered way. The flavor-blind principle, in this respect, helps us to understand how mixing can be produced in such sequential manner. This condition can be applied to the general case of $n$ fermion families where the symmetry properties of the introduced set of Yukawa couplings follow a unique sequential breaking $S_{nL}\otimes S_{nR} \rightarrow S_{(n-1)L}\otimes S_{(n-1)R} \rightarrow \cdots 
	\rightarrow S_{2L}\otimes S_{2R} \rightarrow S_{2A}$. The particular cases of two and three families have been treated. We need to remark that both of them were already studied by other authors but under different considerations.
	Regarding our work, and its main difference with respect to previous ones, we have shown that the last Yukawa matrix that gives mass to the first family should always obey the symmetry properties of the diagonal and antisymmetric subgroup $S_{2A}$. For last, the flavor-blind principle
	has helped us to understand why the four mass ratios parametrization found for the lepton sector~\cite{Hollik:2014jda} equally applies, as in the quark sector, in spite of the very mild hierarchy in the neutrino masses.

\section*{Acknowledgements}
UJSS wants to thank U. Nierste for suggesting the approach of taking the Gatto-Sartori-Tonin relation as a starting point and studying thereof its consequences. Also the author wants to thank A. Aranda, O. G. Miranda, E. Peinado, and 
A. Vicente for a careful reading of the manuscript and their useful comments on it. After submission to the journal the author received many insights and comments from W. G. Hollik; he feels indebted to him.
This work has been supported by the mexican grants CONACyT-166639 and SNI-Mexico.

\appendix

\section{The Gatto-Sartori-Tonin relation in a system with two coupled quantum harmonic oscillators}

\label{App:2qho}
	The most general Hamiltonian describing two coupled quantum harmonic oscillators with 
	\textit{different} frequencies is
	\begin{eqnarray}
		\hat{H} = \frac{\hat{p}_1^2}{2m_1} + \frac{\hat{p}_2^2}{2m_2} + \frac{1}{2}k_1 \hat{x}_1^2 + 
		\frac{1}{2}k_2 \hat{x}_2^2 + 
		\frac{1}{2}k_0 (\hat{x}_2-\hat{x}_1)^2,
	\end{eqnarray}
	where the $k_i$ are real parameters and $\hat{x}_i$ measures the displacement from the equilibrium point.
	Now, in order to decouple them, we first need to make a change of scale but which preserves the corresponding commuting
	relations 
	\begin{eqnarray}
		[\hat{x}_i,\hat{p}_j] = i\hbar \delta_{ij}, \quad\quad [\hat{x}_i,\hat{x}_j] = 0, \quad\quad 
		[\hat{p}_i,\hat{p}_j] = 0.
	\end{eqnarray}
	This change of scale is found to be given by
	\begin{align}
	\hat{p}_1^2 = \sqrt{\frac{m_1}{m_2}} \hat{P}_1^2, \quad\quad\quad
	\hat{x}_1^2 = \sqrt{\frac{m_2}{m_1}} \hat{X}_1^2, \quad\quad\quad
	\hat{p}_2^2 = \sqrt{\frac{m_2}{m_1}} \hat{P}_2^2, \quad\quad\quad
	\hat{x}_2^2 = \sqrt{\frac{m_1}{m_2}} \hat{X}_2^2.
	\end{align}
	The new Hamiltonian becomes
	\begin{eqnarray}
		\hat{H} = \frac{\hat{P}_1^2}{2\mu} + \frac{\hat{P}_2^2}{2\mu} + \frac{1}{2}\mu w_1^2 \hat{X}_1^2 + 
		\frac{1}{2}\mu w_2^2 \hat{X}_2^2 + 
		\frac{1}{2}k_0 \left( \sqrt{\frac{m_2}{m_1}} \hat{X}_1^2 + 
		\sqrt{\frac{m_1}{m_2}} \hat{X}_2^2  + 2 \hat{X}_1 \hat{X}_2 \right),
	\end{eqnarray}
	where $\mu = \sqrt{m_1 m_2}$ is the geometric mass. The system can be brought to diagonal form by means of an
	orthogonal transformation
	\begin{eqnarray}
	 \begin{pmatrix}
		\hat{X}_+ \\ \hat{X}_- 
	\end{pmatrix}	 = \begin{pmatrix}
			\cos\theta \hat{X}_1 -\sin\theta \hat{X}_2 \\
			\sin\theta \hat{X}_1 + \cos\theta \hat{X}_2
		\end{pmatrix},
	\end{eqnarray}
	with the angle of rotation satisfying
	\begin{align}
	\tan 2\theta = \frac{- 2k_0}{\mu(w_2^2 - w_1^2) + k_0\frac{m_1 - m_2}{\mu}}.
	\end{align}
	Thus, the Hamiltonian in terms of the two normal modes now describes two uncoupled harmonic oscillators with equal 
	masses
	\begin{eqnarray}
		\hat{H} = \frac{\hat{P}_+^2}{2\mu} + \frac{\hat{P}_-^2}{2\mu} + \frac{1}{2}\mu w^2_+ \hat{X}_+^2 + 
		\frac{1}{2}\mu w^2_- \hat{X}_-^2,
	\end{eqnarray}
	but two different frequencies
	\begin{eqnarray}
		w_+^2 = w_1^2 \sin^2\theta + w_2^2 \cos^2\theta + \frac{k_0}{\mu}\left( \sqrt[4]{\frac{m_2}{m_1}}\cos\theta + \sqrt[4]{\frac{m_1}{m_2}}\sin\theta \right)^2,\\
		w_-^2 = w_1^2 \cos^2\theta + w_2^2 \sin^2\theta + \frac{k_0}{\mu}\left( \sqrt[4]{\frac{m_1}{m_2}}\cos\theta - \sqrt[4]{\frac{m_2}{m_1}}\sin\theta \right)^2. 
	\end{eqnarray}

	If we ask the system to be in resonance then the initial frequencies need to be equal, $w = w_1 = w_2$. 
	This simplification	
	with physical meaning implies that the rotation angle will satisfy the quadratic equation	
	\begin{align}
		\tan^2\theta + \frac{m_2 - m_1}{\mu} \tan\theta -1 = 0,
	\end{align}
	that is
	\begin{align}
	\tan\theta = \begin{cases}
    -\sqrt{\frac{m_2}{m_1}}\\
    {}\\
    + \sqrt{\frac{m_1}{m_2}}
	  \end{cases},
	\end{align}
	and the new normal frequencies will be
	\begin{align}
	w_+^2 = w^2 + \frac{k_0}{\mu}\left( \sqrt[4]{\frac{m_2}{m_1}}\cos\theta - \sqrt[4]{\frac{m_1}{m_2}}\sin\theta \right)^2 = \begin{cases}
	 w^2, \\
	 w^2 + \frac{k_0}{\mu^2}(m_1+m_2),
	\end{cases}\\ \nonumber{} \\
		w_-^2 = w^2  + \frac{k_0}{\mu}\left( \sqrt[4]{\frac{m_1}{m_2}}\cos\theta + \sqrt[4]{\frac{m_2}{m_1}}\sin\theta \right)^2 = \begin{cases}
		w^2 + \frac{k_0}{\mu^2}(m_1+m_2), \\
		w^2.
		\end{cases}
	\end{align}
	
	Both types of solutions give the same kind of normal frequencies. 
	What we found out from this exercise is that the angle of rotation,
	which brings our system to its normal modes, satisfy the GST relation
	\begin{eqnarray}
		\tan^2 \theta = \frac{m_1}{m_2}.
	\end{eqnarray}
	Hence the GST relation originates, in this case, by uncoupling and 
	demanding resonance (same frequencies)
	to the two coupled quantum harmonic oscillators with different masses.

\section{Application of the flavor-blind principle to the $n$ family case}
\label{app:nFamiliesCase}

Let us consider $n$ fermion families, that is $n$ fermion fields 
with universal gauge couplings. Their kinetic terms when putted 
together span an $n$-dimensional family space. Any unitary
transformation acting in the $n$ family vector 
of each gauge representation will leave the kinetic
terms invariant. This group commonly known as the flavor group is 
given as\footnote{For the sake of clarity we choose to work only with the Dirac case.}
\begin{eqnarray}
	{\cal G}_F = U(n)_L^Q \otimes U(n)_R^u \otimes U(n)_R^d
	\otimes 
	U(n)_L^\ell \otimes U(n)_R^e \otimes U(n)_R^\nu.
\end{eqnarray}

Application of the flavor-blind principle tells us how to
introduce the Yukawa couplings. For a generic fermion species 
the first matrix term cannot distinguish among the $n$ fermion
fields. Thus,
\begin{eqnarray} \label{eq:n-permutation}
	{\cal Y}^{1\leftrightarrow 2 \leftrightarrow 3 \leftrightarrow
	 \cdots \leftrightarrow (n-1) \leftrightarrow n} = y_n \begin{pmatrix}
	 1 & 1 & 1 & \cdots & 1 \\
	 1 & 1 & 1 & \cdots & 1 \\
	 1 & 1 & 1 & \cdots & 1 \\
	\vdots & \vdots & \vdots & \vdots & \vdots \\
	1 & 1 & 1 & \cdots & 1 	 
	 \end{pmatrix}.
\end{eqnarray} 
This basis is known as the democratic basis. The symmetry properties of this matrix are described by the group $S_{nL}\otimes S_{nR}$. Usage of the weak basis transformation
\begin{eqnarray} \label{WBT-Sn}
	O_n = \begin{pmatrix}
\frac{1}{\sqrt{2}} & \frac{1}{\sqrt{3\cdot 2}} & \frac{1}{\sqrt{4\cdot 3}} & \cdots &
\frac{1}{\sqrt{n(n-1)}} &\frac{1}{\sqrt{n}} \\
-\frac{1}{\sqrt{2}} & \frac{1}{\sqrt{3\cdot 2}} & \frac{1}{\sqrt{4\cdot 3}} & \cdots &
\frac{1}{\sqrt{n(n-1)}} & \frac{1}{\sqrt{n}} \\
 0 & -\frac{2}{\sqrt{3\cdot 2}} & \frac{1}{\sqrt{4\cdot 3}} & \cdots &
\frac{1}{\sqrt{n(n-1)}} & \frac{1}{\sqrt{n}} \\
0 & 0 &  -\frac{3}{\sqrt{4\cdot 3}} & \cdots &
\frac{1}{\sqrt{n(n-1)}} & \frac{1}{\sqrt{n}} \\
\vdots & \vdots & \vdots & \cdots & \vdots & \vdots \\
0 & 0 & 0 & \cdots & \frac{1}{\sqrt{n(n-1)}} & \frac{1}{\sqrt{n}}\\
0 & 0 & 0 & \cdots & -\frac{n-1}{\sqrt{n(n-1)}} & \frac{1}{\sqrt{n}}
	\end{pmatrix},
\end{eqnarray}
brings all Yukawa matrices to the form
\begin{eqnarray} \label{eq:nnelelement}
	\widetilde{\cal Y}^{1\leftrightarrow 2 \leftrightarrow 3 \leftrightarrow
	 \cdots \leftrightarrow (n-1) \leftrightarrow n} = n y_n \begin{pmatrix}
	 0 & 0 & 0 & \cdots & 0 \\
	 0 & 0 & 0 & \cdots & 0 \\
	 0 & 0 & 0 & \cdots & 0 \\
	\vdots & \vdots & \vdots & \vdots & \vdots \\
	0 & 0  & 0 & \cdots & 1 	 
	 \end{pmatrix}.
\end{eqnarray} 
From Eq.~(\ref{eq:n-permutation}) we already know we will have a single massive family, as it is a rank one matrix. 
In the new basis, or heavy basis, the mass term is $m_n = n y_n v$ with $v\equiv \braket{H}$ the nonvanishing vacuum expectation value of the neutral component of the Higgs field. Meanwhile, ${\cal G}_F$ has broken to $U(n-1)^6$.

Again, the flavor-blind principle helps us to introduce the new Yukawa couplings. In the democratic basis we have
\begin{eqnarray}
	{\cal Y}^{1\leftrightarrow 2 \leftrightarrow 3 \leftrightarrow
	 \cdots \leftrightarrow (n-1)} = \begin{pmatrix}
	 	\beta_{n-1} & \beta_{n-1} & \cdots & \beta_{n-1} & \alpha_{n-1} \\
	 	\beta_{n-1} & \beta_{n-1} & \cdots & \beta_{n-1} & \alpha_{n-1} \\
	 	\vdots & \vdots & \cdots & \vdots & \vdots \\
	 	\beta_{n-1} & \beta_{n-1} & \cdots & \beta_{n-1} & \alpha_{n-1} \\
	 	\alpha'_{n-1} & \alpha'_{n-1} &\cdots & \alpha'_{n-1} & \gamma_{n-1}
	 \end{pmatrix}.
\end{eqnarray}
This is a rank two matrix. 
The arbitrary parameters $\alpha_{n-1}$, $\beta_{n-1}$, $\alpha'_{n-1}$, and
$\gamma_{n-1}$ are in general complex parameters; all of them depending in $m_{n-1}/v$ and $m_n/v$. In the latter case
as an homogeneous function of first order. Whenever $m_{n-1} = 0$ all of them are also equal to zero. 
The $(n-1)$-dimensional submatrix in the upper left corner suggests the possibility of
extracting a similar matrix as in Eq.~(\ref{eq:n-permutation}). This is only telling us that, after the weak basis transformation of Eq.~(\ref{WBT-Sn}), 
the $(n,n)$ element in Eq.~(\ref{eq:nnelelement}) will gain a contribution proportional to $m_{n-1}$.
The heavy basis shows explicitly how $U(n-1)^6$ has broken to $U(n-2)^6$ with only four non-zero
matrix elements
\begin{eqnarray}
\widetilde{\cal Y}^{1\leftrightarrow 2 \leftrightarrow 3 \leftrightarrow
	 \cdots \leftrightarrow (n-1)}_{n-1,n-1} = 
	 \frac{n-1}{n}[\beta_{n-1} + \gamma_{n-1} - \alpha_{n-1}-\alpha'_{n-1} ],  \\
\widetilde{\cal Y}^{1\leftrightarrow 2 \leftrightarrow 3 \leftrightarrow
	 \cdots \leftrightarrow (n-1)}_{n-1,n} = 
	  \frac{\sqrt{n-1}}{n}[(n-1)(\beta_{n-1} - \alpha'_{n-1}) + \alpha_{n-1} - \gamma_{n-1}], \\
\widetilde{\cal Y}^{1\leftrightarrow 2 \leftrightarrow 3 \leftrightarrow
	 \cdots \leftrightarrow (n-1)}_{n,n-1} =  
	 \frac{\sqrt{n-1}}{n}[(n-1)(\beta_{n-1} - \alpha_{n-1}) + \alpha'_{n-1} - \gamma_{n-1}],  \\
\widetilde{\cal Y}^{1\leftrightarrow 2 \leftrightarrow 3 \leftrightarrow
	 \cdots \leftrightarrow (n-1)}_{n,n} = 
	  \frac{1}{n}\{ \gamma_{n-1} + (n-1)[(n-1)\beta_{n-1}+\alpha_{n-1}+\alpha'_{n-1}] \}.
\end{eqnarray}
That is, $n-2$ families
are massless. The flavor-blind principle tells us that the new Yukawa couplings to be introduced
in the democratic basis should be
\begin{eqnarray}
	{\cal Y}^{1\leftrightarrow 2 \leftrightarrow 3 \leftrightarrow
	 \cdots \leftrightarrow (n-2)} = \begin{pmatrix}
	 	\beta_{n-2} & \beta_{n-2} & \cdots & \beta_{n-2} & \omega_{n-2} &\alpha_{n-2} \\
	 	\beta_{n-2} & \beta_{n-2} & \cdots & \beta_{n-2} &\omega_{n-2} &  \alpha_{n-2} \\
	 	\vdots & \vdots & \cdots & \vdots & \vdots & \vdots \\
	 	\beta_{n-2} & \beta_{n-2} & \cdots & \beta_{n-2}&\omega_{n-2} & \alpha_{n-2} \\
	 	\omega'_{n-2} & \omega'_{n-2} & \cdots & \omega'_{n-2} &\rho_{n-2} & \lambda_{n-2} \\
	 	\alpha'_{n-2} & \alpha'_{n-2} &\cdots & \alpha'_{n-2} & \lambda'_{n-2} & \gamma_{n-2}
	 \end{pmatrix}.
\end{eqnarray}
which after the weak basis transformation the only non-zero elements are those conforming
the $3\times 3$ submatrix
\begin{eqnarray}
\widetilde{\cal Y}^{1\leftrightarrow 2 \leftrightarrow 3 \leftrightarrow
	 \cdots \leftrightarrow (n-2)}_{n-2,n-2} = \frac{n-2}{n-1}[\beta_{n-2} + \rho_{n-2}
	 - \omega_{n-2} -\omega'_{n-2}],  \\
\widetilde{\cal Y}^{1\leftrightarrow 2 \leftrightarrow 3 \leftrightarrow
	 \cdots \leftrightarrow (n-2)}_{n-2,n-1} =	 \sqrt{\frac{n-2}{n-1}} 
	 [(n-2)(\beta_{n-2}-\omega'_{n-2}) -(n-1)( \alpha_{n-2} - \lambda_{n-2})  + \omega_{n-2} - \rho_{n-2}], \\
	\widetilde{\cal Y}^{1\leftrightarrow 2 \leftrightarrow 3 \leftrightarrow
	 \cdots \leftrightarrow (n-2)}_{n-2,n} = 
	 \sqrt{\frac{n-2}{n(n-1)}}[(n-2)(\beta_{n-2} - \omega'_{n-2}) + \alpha_{n-2} - \lambda_{n-2} 
	 + \omega_{n-2} - \rho_{n-2}], \\
	 \widetilde{\cal Y}^{1\leftrightarrow 2 \leftrightarrow 3 \leftrightarrow
	 \cdots \leftrightarrow (n-2)}_{n-1,n-2} =	 \sqrt{\frac{n-2}{n-1}} 
	 [(n-2)(\beta_{n-2}-\omega_{n-2}) -(n-1)( \alpha'_{n-2} - \lambda'_{n-2})  + \omega'_{n-2} - \rho_{n-2}], \\ \nonumber
	 \widetilde{\cal Y}^{1\leftrightarrow 2 \leftrightarrow 3 \leftrightarrow
	 \cdots \leftrightarrow (n-2)}_{n-1,n-1} =	 \frac{1}{n(n-1)}\{(n-2)[(n-2)\beta_{n-2}
	 + \omega'_{n-2} + \omega_{n-2} + \alpha_{n-2} + \alpha'_{n-2}] + \\ 
	 (n-1)[(n-1)\gamma_{n-2} - \lambda'_{n-2} - \lambda_{n-2}] + \rho_{n-2}, \} 
\end{eqnarray}
\begin{eqnarray}\nonumber
	 \widetilde{\cal Y}^{1\leftrightarrow 2 \leftrightarrow 3 \leftrightarrow
	 \cdots \leftrightarrow (n-2)}_{n-1,n} = \frac{1}{n\sqrt{n-1}}\{ (n-2)[(n-2)\beta_{n-2} 
	 + \omega'_{n-2} + \omega_{n-2} -(n-1)\alpha'_{n-2} + \alpha_{n-2}] 	 \\ 
	 -(n-1)(\gamma_{n-2} + \lambda'_{n-2}) + \rho_{n-2} + \lambda_{n-2}
	 \},  \\
	 \widetilde{\cal Y}^{1\leftrightarrow 2 \leftrightarrow 3 \leftrightarrow
	 \cdots \leftrightarrow (n-2)}_{n,n-2} =  
	  \sqrt{\frac{n-2}{n(n-1)}}[(n-2)(\beta_{n-2} - \omega_{n-2}) + \alpha'_{n-2} - \lambda'_{n-2} 
	 + \omega'_{n-2} - \rho_{n-2}], \\ \nonumber
	 \widetilde{\cal Y}^{1\leftrightarrow 2 \leftrightarrow 3 \leftrightarrow
	 \cdots \leftrightarrow (n-2)}_{n,n-1} =  \frac{1}{n\sqrt{n-1}}\{ (n-2)[(n-2)\beta_{n-2} 
	 + \omega_{n-2} + \omega'_{n-2} -(n-1)\alpha_{n-2} + \alpha'_{n-2}] 	 \\ 
	 -(n-1)(\gamma_{n-2} + \lambda_{n-2}) + \rho_{n-2} + \lambda'_{n-2}
	 \}, \\ \nonumber
	  \widetilde{\cal Y}^{1\leftrightarrow 2 \leftrightarrow 3 \leftrightarrow
	 \cdots \leftrightarrow (n-2)}_{n,n} = \frac{1}{n}\{(n-2)[(n-2)\beta_{n-2} + \omega'_{n-2} + \omega_{n-2} + \alpha'_{n-2} + \alpha_{n-2}] + \\
	  \rho_{n-2} + \gamma_{n-2} 	 + \lambda'_{n-2} + \lambda_{n-2} \}. 
\end{eqnarray}
Inclusion of further terms to give masses to the $n-2$ fermions left will follow the same logic 
as given by the flavor-blind principle.

For last, let us discuss why the last step of symmetry breaking is given by $S_{2L} \otimes S_{2R} \rightarrow S_{2A}$ instead of $S_{2L} \otimes S_{2R} \rightarrow S_{2S}$. In the former case,
the Yukawa couplings are given by
\begin{eqnarray}
	{\cal Y}^A = \begin{pmatrix}
		\beta_1 & \xi_1 & \lambda_1 & \cdots & \alpha_1 \\
		-\xi_1 & -\beta_1 & -\lambda_1 & \cdots & -\alpha_1 \\
		\varsigma_1 & -\varsigma_1 & 0 & \cdots & c_1 \\
		\vdots & \vdots &\vdots & \vdots & \vdots \\
		\mu_1 & -\mu_1 & c'_1 &\cdots & 0
	\end{pmatrix},
\end{eqnarray}
whereas for the latter
\begin{eqnarray}
	{\cal Y}^S = \begin{pmatrix}
	   \beta_1 & \xi_1 & \lambda_1 & \cdots & \alpha_1 \\
		\xi_1 & \beta_1 & \lambda_1 & \cdots & \alpha_1 \\
		\varsigma_1 & \varsigma_1 & a_1 & \cdots & c_1 \\
		\vdots & \vdots &\vdots & \vdots & \vdots \\
		\mu_1 & \mu_1 & c'_1 &\cdots & z_1
	\end{pmatrix}.
\end{eqnarray}
A straightforward calculation shows
 that after the corresponding electroweak basis transformation one gets 
the generic forms\footnote{Only for $n=3$ the generic form of $\widetilde{\cal Y}^A$ in Eq.~(\ref{nf-yA}) acquires the form written in Eq.~(\ref{3fermion-yA}).}
\begin{eqnarray}\label{nf-yA}
	\widetilde{\cal Y}^A \sim \begin{pmatrix}
	0 & \times & \times & \cdots & \times \\
	\times & 0 & \times &\cdots & \times \\
	\times & \times & \times & \cdots & \times \\
	\vdots & \vdots &\vdots & \vdots & \vdots \\
	\times & \times & \times & \cdots & \times
	\end{pmatrix},
\end{eqnarray}
and
\begin{eqnarray}
	\widetilde{\cal Y}^S \sim \begin{pmatrix}
		\times & 0 & 0 &\cdots & 0 \\
		0 & \times & \times & \cdots & \times \\
		0 & \times & \times &\cdots & \times \\
		\vdots & \vdots  & \vdots & \vdots & \vdots \\
		0 & \times & \times & \cdots & \times
	\end{pmatrix}.
\end{eqnarray}
From this picture it becomes clear why $S_{2L} \otimes S_{2R} \rightarrow S_{2A}$  is the correct
symmetry breaking step as it is the only one which implies mixing between the first family with the other fermion families.

\bibliographystyle{utcaps}	
\bibliography{GST}

\end{document}